\documentclass[a4paper,twocolumn,tabularx]{article}
\usepackage{graphicx}
\begin{document}
\newcommand{\gtrsim}{\stackrel{>}{\sim}}
\newcommand{\lesssim}{\stackrel{<}{\sim}}

\title{Coulomb blockade in one-dimensional arrays of high conductance
tunnel junctions}

\author{\small Sh. Farhangfar, R.S. Poikolainen, J.P. Pekola \\
\small\it Department of Physics, University of Jyv{\"a}skyl{\"a},
P.O. Box 35 (Y5), FIN-40351 Jyv{\"a}skyl{\"a}, Finland\\ \small
D.S. Golubev and A.D. Zaikin \\ \small\it Forschungszentrum
Karlsruhe, Institut f{\"u}r Nanotechnologie, D-76021 Karlsruhe,
Germany; and
\\ \small\it I.E. Tamm Department of Theoretical Physics, P.N. Lebedev
Physics Institute,
\\ \small\it Leninskii pr. 53, 117924 Moscow, Russia}

\date{\today}
\maketitle
\begin{abstract}
Properties of one-dimensional (1D) arrays of low Ohmic
tunnel junctions ({\it i.e.} junctions with resistances comparable to,
or less than, the quantum resistance
$R_{\rm q}\equiv h/e^2\approx 25.8$ k$\Omega$)
have been studied experimentally and theoretically. Our experimental data
demonstrate that -- in agreement with previous results on
single- and double-junction systems -- Coulomb blockade effects survive
even in the strong tunneling regime and are still clearly visible for
junction resistances as low as 1 k$\Omega$.
We have developed a quasiclassical theory of
electron transport in junction arrays in the strong tunneling regime.
Good agreement between the predictions of this theory and the experimental
data has been observed.
We also show that, due to both heating effects {\it and} a relatively
large correction to the linear relation between the half-width of the
conductance dip around zero bias voltage, $V_{1/2}$, and the measured electronic temperature,
such arrays  are inferior to those conventionally used in the Coulomb
Blockade Thermometry (CBT). Still, the desired correction to the half-width, $\Delta V_{1/2}$, can be
determined rather easily and it is proportional to the magnitude
of the conductance dip around zero bias voltage, $\Delta G$. The constant of proportionality is a function of  the
ratio of the junction and quantum resistances, $R/R_{\rm q}$, and it is a {\sl pure} strong tunneling effect.
\end{abstract}

\section{Introduction}

The effect of  Coulomb blockade in 1D arrays of normal tunnel
junctions can be used for absolute thermometry \cite{exp1}-\cite{exp2}.
The properties of such arrays have been extensively investigated both
experimentally and theoretically \cite{exp1}-\cite{exp3}. In all these
works  arrays of high Ohmic junctions with the junction resistances,
$R_j$,
higher than or of the order of the quantum resistance $R_{\rm q}=h/e^2\approx 25.8$
k$\Omega$ have been studied. In this limit
a theoretical description of the Coulomb blockade
is well developed \cite{AL} and it has been successfully applied
\cite{exp1}-\cite{exp2} to explain experimental findings in the high
temperature regime,
${k_B}T> E_C\equiv {e^2/2C}$,
where $C$ is the capacitance of a single junction.

From the practical point of view, large ($N \gg 1$) 1D arrays used as thermometers
are advantageous over the smaller ones, because of  higher
accuracy of temperature measurements. At the same time, increasing
the number of junctions in the array obviously yields an increase
of its total resistance $R_{\rm tot} \propto N$. Since in practice
it is desirable to avoid very large values of $R_{\rm tot}$, it
appears natural to relax the condition $R_j \gg R_{\rm q}$ and
consider 1D arrays of relatively highly conducting tunnel
junctions with $R_j$ of the order of $R_{\rm q}$ or smaller. A natural way to avoid this is the
parallel connection of several 1D arrays, and this is used extensively in Coulomb blockade
thermometry (CBT). To avoid a large total number of junctions, a more straightforward
solution would be to decrease the resistance of each individual junction.
On the other hand, heating effects turn out to be much more pronounced
for highly conducting junction arrays
\cite{exp4}. Hence, from this point of view, it is better {\it not}
to decrease $R_j$ down to  very low values.

The above considerations motivated us to  investigate the
interplay between Coulomb blockade and strong tunneling
effects in 1D arrays of normal metallic tunnel junctions. Do Coulomb
blockade effects survive if the junction resistance becomes smaller than
the resistance quantum? Both theory \cite{PZ}-\cite{G}
and experiment \cite{Chalmers}-\cite{Helsinki} give a clear
positive answer to this question. An adequate theoretical
approach which enables one to study electron transport in the
strong tunneling regime is well established \cite{AES,GZ1}.
This so-called quasiclassical Langevin equation technique allows to proceed
analytically and remains accurate at not very low temperatures and/or voltages
\cite{GZ1}:
\begin{equation}
{\rm max}[k_BT,eV] \gtrsim (\hbar/R_jC)\exp (-R_{\rm q}/2R_j) \label{1}
\end{equation}
for $R_j \lesssim R_{\rm q}$, and ${\rm max}[k_BT,eV] \gtrsim e^2/2C$
otherwise. The condition (\ref{1}) implies that in the strong
tunneling regime $R_j \ll R_{\rm q}$, which is of a primary interest for
us here, the technique \cite{AES,GZ1} covers practically all
experimentally accessible values of temperature and bias voltage.
In several previous publications \cite{GZ1}-\cite{G}, the Langevin
equation technique was applied to analyze the Coulomb blockade and
strong tunneling effects in single junctions and SET transistors,
where the $I-V$ curves were derived at arbitrary
tunneling strength. The results of this theoretical analysis
turned out to be in  good agreement with available experimental data
\cite{Chalmers,Saclay,Helsinki}.

The structure of the paper is as follows: In Section 2 we develop a
theoretical analysis of Coulomb blockade effects in 1D tunnel junction arrays
in the strong tunneling regime. In Section 3 we present experimental
results obtained for the arrays with resistances in the range $\sim 1-23$ k$\Omega$
and compare these results with our theoretical predictions.
Our main conclusions are summarized in Section 4.

\begin{figure}
\begin{center}
\includegraphics[height=4.0cm]{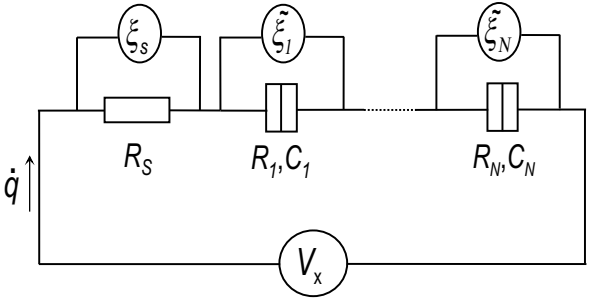}
\end{center}
\caption{A 1D array of normal metal tunnel junctions. }
\end{figure}

\section{Theory}

In order to theoretically study the
behavior of low resistance tunnel junction
arrays we are going to use the technique of quasiclassical Langevin
equation
developed in Refs. \cite{AES,GZ1}.
In this section we generalize the approach presented in \cite{GZ1}
to the case of 1D arrays of tunnel junctions.

We consider an array of $N$ normal metal tunnel junctions in series (Fig. 1).
The system can be described by the following Langevin equations
\cite{GZ1}-\cite{G}:
\begin{eqnarray}
&C_j& \frac{\hbar\ddot\varphi_j}{2e} +
\frac{1}{R_j}\frac{\hbar\dot\varphi_j}{2e}
 = \dot q +  \tilde\xi_j, {\hskip7mm} j=1,\ldots,N;
\nonumber \\
&V_{\rm x}&=\dot q R_{\rm S}+ \sum\limits_{j=1}^N\frac{\hbar\dot\varphi}{2e}
- \xi_SR_{\rm S}.
\label{balance}
\end{eqnarray}
Here $ \varphi_j(t)\equiv\frac{2e}{\hbar}\int_{0}^{t}dt^{\prime}
V_j(t^\prime)$ is the effective phase
and $R_{\rm S}$ is the resistance of the electromagnetic environment. The shot
noise of the $j$-th junction depends on $\varphi_j$ as \cite{AES,GZ1}:
\begin{equation}
\tilde\xi_j=\xi_{1j}\cos(\frac{\varphi_j}{2})+\xi_{2j}\sin(\frac{\varphi_j}{2}).
\label{tilde}
\end{equation}
Here $\xi_{jk}$ are Gaussian stochastic variables with the following
pair correlators:
\begin{eqnarray}
&\langle\xi_{1j}(t_1)\xi_{1j}(t_2)\rangle&=
\langle\xi_{2j}(t_1)\xi_{2j}(t_2)\rangle=
\frac{G(t_1-t_2)}{R_j},
\nonumber\\
&\langle\xi_{1j}(t_1)\xi_{2j}(t_2)\rangle&=0, \;\;
\langle\xi_S(t_1)\xi_S(t_2)\rangle=
\frac{G(t_1-t_2)}{R_S}.
\label{noises}
\end{eqnarray}
In Eqs. (\ref{noises}) we have defined
\begin{equation}
G(t)\equiv\int\limits _{-\infty}^{+\infty}
\frac{d\omega}{2\pi} \hbar \omega \coth(\frac{\hbar\omega}{2k_BT})
{\rm e}^{i\omega t}
= -\frac{1}{\pi\hbar}{\cal P}
\frac{(\pi k_B T)^2} {\sinh^2(\frac{\pi k_B Tt}{\hbar})},
\label{G}
\end{equation}
where ${\cal P}$ stands for the principal value.
There exist no correlations between the noise terms from different
junctions:
$\langle\xi_{1(2)i}(t_1)\xi_{1(2)j}\rangle= 0$ for $i \neq j$.

Averaging Eqs. (\ref{balance}) over the noise realizations
(we will denote this average by angular brackets),
we obtain the general expression for the current in the array:
\begin{eqnarray}
I(V_x)&=&\langle\dot
q\rangle=\frac{V}{R_\Sigma}-\sum\limits_{j=1}^N\frac{R_j}{R_\Sigma}
\langle\tilde\xi_j\rangle
\nonumber\\
V(V_x)&=&\bigg\langle \sum\limits_{j=1}^N
\frac{\hbar\dot\varphi_j}{2e}\bigg\rangle,
\label{IV}
\end{eqnarray}
where $R_\Sigma\equiv\sum_{j=1}^{N} R_j$ is the total resistance of the
array.
These two equations define the $I-V$ curve of the chain.

The problem, now, reduces to the evaluation of the noise averages
$\langle\tilde\xi_j\rangle$. In order to do this, we first exclude the
current
$\dot q$ from the equations (\ref{balance}) and get the following ones
for the phases $\varphi_j$:
\begin{equation}
C_j\frac{\hbar\ddot\varphi_j}{2e} +
\frac{1}{R_j}\frac{\hbar\dot\varphi_j}{2e}+
\frac{1}{R_S}\sum\limits_{k=1}^N\frac{\hbar\dot\varphi_k}{2e}=
\frac{V_x}{R_S}+\xi_S+\tilde\xi_j.
\label{eq1}
\end{equation}
Then we define the small deviations of the phase and the shot noise from
their
average values   $\delta\varphi_j {\equiv}{\hskip1mm}
\varphi_j-\langle\varphi_j\rangle$ and
$\delta\tilde\xi_j {\equiv}{\hskip1mm}
\tilde\xi_j-\langle\tilde\xi_j\rangle$, respectively. They
obey the following equations
\begin{equation}
C_j\frac{\hbar\delta\ddot\varphi_j}{2e} +
\frac{1}{R_j}\frac{\hbar\delta\dot\varphi_j}{2e}+
\frac{1}{R_S}\sum\limits_{k=1}^N\frac{\hbar\delta\dot\varphi_k}{2e}=
\xi_S+\delta\tilde\xi_j.
\label{fluct}
\end{equation}
Applying the Fourier transformation and solving the corresponding
equations, we find
\begin{equation}
\left.\frac{\hbar\delta\dot\varphi_j}{2e}\right|_{\omega}=
Z_j(\omega)\delta\tilde\xi_{j,\omega}+\sum\limits_{k\not=j}a_{jk}(\omega)
\delta\tilde\xi_{k,\omega},
\label{deltaphij}
\end{equation}
where $Z_j(\omega)$ is the total impedance seen by the $j$-th junction
and
the functions $a_{jk}(\omega)$ describe the mutual influence of the
junctions on each other. Equation (\ref{deltaphij}) applies to
any array of tunnel junctions of any dimensionality. In the case of 1D
arrays from Eq. (\ref{fluct}) we find
\begin{equation}
Z_j(\omega)=\frac{R_j}{1-i\omega
R_jC_j}\frac{R_S+\sum\limits_{k\not=j}\frac{R_k}{1-i\omega R_kC_k}}
{R_S+\sum\limits_{k}\frac{R_k}{1-i\omega R_kC_k}}.
\label{Zj}
\end{equation}
Here we do not present the expressions for the functions
$a_{jk}(\omega)$
because, as we will see below, the contribution of corresponding terms
turns out to vanish. We find
\begin{eqnarray}
\delta\varphi_j&=&\frac{2e}{\hbar}\int\limits_{-\infty}^t dt'
K_j(t-t')\delta\tilde\xi_j(t')
\nonumber\\
&&+
\sum\limits_{k\not=j}\frac{2e}{\hbar}
\int\limits_{-\infty}^t dt'
A_{jk}(t-t')\delta\tilde\xi_{k}(t'),
\label{dphi1}
\end{eqnarray}
where the response function $K_j(t)$ is defined as
\begin{equation}
K_j(t)=\int\limits_{-\infty}^{+\infty}\frac{d\omega}{2\pi}\;
\frac{Z_j(\omega)}{-i\omega+0}\;{\rm e}^{-i\omega t},
\label{K}
\end{equation}
and $A_{jk}(t)$ are defined analogously.

It is important to emphasize that Eqs. (\ref{deltaphij}) and
(\ref{dphi1}) are not the explicit solutions for the phase, but
the integral equations. The variables $\delta\tilde\xi_j$ on the
right hand side of these equations depend on $\delta\varphi_j$
through the $\sin(\varphi_{j}/2)$ and $\cos(\varphi_{j}/2)$ terms in the
shot noise (\ref{tilde}). These integral  equations can be solved
by iteration. Here we restrict ourselves to the first iteration
and put $\delta\varphi_j=0$ in the right hand side of Eq.
(\ref{dphi1}).

The next step is to evaluate the average values
$\langle\tilde\xi_j\rangle,$ which  enter the expression for the
current (\ref{IV}). We make the following approximation:
\begin{eqnarray}
\langle\tilde\xi_j\rangle=
\bigg\langle\xi_{1j}\cos\left[\frac{eV_jt}{\hbar}+\frac{\delta\varphi_j}{2}\right]
+
\xi_{2j}\sin\left[\frac{eV_jt}{\hbar}+\frac{\delta\varphi_j}{2}\right]\bigg\rangle
\nonumber\\
\simeq
\bigg\langle\left[\xi_{2j}\cos\frac{eV_jt}{\hbar}-\xi_{1j}\sin\frac{eV_jt}{\hbar}\right]
\frac{\delta\varphi_j}{2}\bigg\rangle.
\;\;\;\;\;\;\;\;\;\;\;\;\;\;\;\;\;
\nonumber
\end{eqnarray}
Here $V_j\equiv\left\langle\frac{\hbar\dot\varphi_j}{2e}\right\rangle$
is the average voltage
on the $j$-th junction. Making use of Eq. (\ref{dphi1}) we get
\begin{equation}
\langle\tilde\xi_j\rangle=-\frac{2e}{\hbar R_j}\int\limits_0^\infty dt\;
G(t)K_j(t)\sin(\frac{eV_jt}{\hbar}).
\label{xi}
\end{equation}
Here we note that the terms containing the kernels $A_{jk}(t)$ do not
contribute to the result because there exists no correlation between the
noise
on different junctions. Now the current is expressed as follows:
\begin{equation}
I=\frac{V}{R_\Sigma}+\frac{2e}{\hbar R_\Sigma}
\int\limits_0^{\infty}dt\; G(t)\bigg[\sum\limits_{j=1}^N
K_j(t)\sin(\frac{eV_jt}{\hbar})\bigg].
\label{IV1}
\end{equation}

Let us first assume that all the junctions in the chain are
identical, {\it i.e.} they have the same resistance $R\equiv R_j$
and the same capacitance $C\equiv C_j$. Then we get
\begin{equation}
Z_j(\omega)= \frac{R_S+(N-1)\frac{R}{1-i\omega
RC}}{R_S\left(\frac{1}{R}-i\omega C\right)+N},
\label{Z}
\end{equation}
\begin{eqnarray}
K_j(t)&=&\frac{N-1}{N}R\left(1-{\rm e}^{-t/RC}\right)
\nonumber\\
&&+\frac{R_SR}{N(R_S+NR)}
\left(1-{\rm e}^{-\frac{R_S+NR}{R_SRC}t}\right).
\label{K1}
\end{eqnarray}
The current (\ref{IV1}), then, can be found exactly:
\begin{equation}
I=\frac{Tv}{eR}-\frac{eT}{\pi\hbar}\left[\frac{N-1}{N}F(v,u)+
\frac{F(v,u_S)}{N(1+N\frac{R}{R_S})}\right].
\end{equation}
Here $v\equiv eV/Nk_BT$, $u\equiv \hbar/2\pi k_BTRC$, $u_S\equiv
u(1+NR/R_S)$, and
\begin{eqnarray}
F(v,u)&\equiv&v\left[{\rm Re}\Psi\left(1+u-i\frac{v}{2\pi}\right)
-{\rm Re}\Psi\left(1-i\frac{v}{2\pi}\right)\right]
\nonumber\\
&&
-2\pi u\; {\rm Im}\Psi\left(1+u-i\frac{v}{2\pi}\right).
\label{F}
\end{eqnarray}
In the limit $T \to 0$ the final result of the above expressions reduces
to
that of our previous analysis \cite{GZ2}.

The differential conductance is given by the following equation
\begin{eqnarray}
R_\Sigma\frac{dI}{dV}&=&1-\frac{e^2R}{\pi\hbar}\bigg[
\frac{N-1}{N}\frac{\partial F(v,u)}{\partial v}
\nonumber\\
&& +
\frac{1}{N(1+N\frac{R}{R_S})}\frac{\partial F(v,u_S)}{\partial v}
\bigg].
\label{dIdV}
\end{eqnarray}

Now let us put $R_{\rm S}=0$ and consider the high temperature limit $u\ll 1$.

Then, in the first order in $u$, we find
\begin{eqnarray}
I&=&\frac{k_BTv}{eR}-\frac{N-1}{N}\frac{ek_BT}{\pi\hbar}u
\left[v\;{\rm Re}\Psi'\left(1-i\frac{v}{2\pi}\right)\right.
\nonumber\\
&&\left. -2\pi\;
{\rm Im}\Psi\left(1-i\frac{v}{2\pi}\right) \right]
\nonumber\\
&=&\frac{k_BTv}{eR}-\frac{N-1}{2N}\frac{e}{RC}\left[
\coth\frac{v}{2}-\frac{v}{2\sinh^2\frac{v}{2}}\right]
\label{IhighT}
\end{eqnarray}
and
\begin{equation}
R_\Sigma\frac{dI}{dV}=1-\frac{N-1}{N}\frac{e^2}{Ck_BT}\frac{v\sinh
v-4\sinh^2\frac{v}{2}}{8\sinh^4\frac{v}{2}}.
\label{dIdVhighT}
\end{equation}
This result exactly coincides with that found for the high Ohmic
junctions
\cite{exp1,exp2}.

Equation (\ref{dIdVhighT}) is basic for the Coulomb blockade
thermometry. The half-width of the dip in the $dI/dV$ is related to
the temperature as $V_{1/2,0}=5.439 Nk_BT/e$ \cite{exp1}. This
relation is proven to be very accurate for the high Ohmic arrays.
To estimate its accuracy in the strong tunneling limit, we use the
expression (\ref{dIdV}), put $R_S=0$ and numerically solve the
following equation
\begin{equation}
\frac{\partial F(v_{1/2}/2,u)}{\partial v}=\frac{1}{2}\frac{\partial
F(0,u)}{\partial v}.
\label{v1/2}
\end{equation}
Here $v_{1/2}$ is the normalized half-width, $v_{1/2}\equiv
V_{1/2}/V_{1/2,0}$. The solution of this equation is plotted in Fig. 2.

\begin{figure}
\begin{center}
\includegraphics[height=6.5cm]{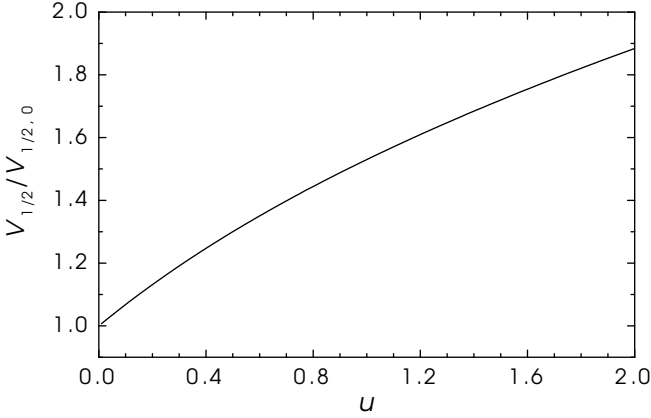}
\end{center}
\caption{The normalized half-width of the conductance dip at zero bias
voltage as a function of (inverse)
temperature $u\equiv{\hbar}/[2\pi {k_B}TRC]$. At higher temperatures the
curve is
almost linear in $u$ [cf. Eq. (22) in the text].}
\end{figure}
{\vskip6mm}
In the limit of high temperatures (small $u$) we find
\begin{eqnarray}
v_{1/2}= 1+0.704 u - 0.24 u^2 + \cdots
\;\;\;\;\;\;\;\;\;\;\;\;\;\;\;\;\;\;
\nonumber\\
=1 + 0.112 \frac{\hbar}{k_BTRC}
-0.006 \frac{\hbar^2}{(k_BTRC)^2}+\cdots.
\label{Vhalfexact}
\end{eqnarray}

The zero bias conductance can be obtained from  Eq.
(\ref{dIdV}):
\begin{eqnarray}
\frac{G_0}{G_\Sigma}=1-\frac{N-1}{N}\frac{e^2R}{\pi\hbar}
\bigg\{\Psi(1+u)+\gamma
+u\,\Psi'(1+u)\bigg\}.
\label{G0}
\end{eqnarray}
Here $G_\Sigma\equiv 1/R_\Sigma$, and $\gamma$ is the Euler's constant.
Note that the result (\ref{G0}) is equivalent to the one recently
derived within the framework of a  linear response
approach  based on the Kubo formula \cite{Freiburg}.
At high temperatures (small $u$) we find from (\ref{G0})
\begin{eqnarray}
R_{\Sigma} G_0=1-\frac{N-1}{N}\frac{e^2R}{\pi\hbar} \left\{
\frac{\pi^2}{3}u - 3\zeta(3) u^2 + \cdots \right\}  \nonumber\\  =
1-\frac{N-1}{N}\frac{e^2}{6Ck_BT}\bigg\{1-0.17446\frac{\hbar}{k_BTRC}+\cdots\bigg\}.
\label{Rsigma}
\end{eqnarray}

Now we will study the effect of the junction asymmetry on the
properties of the thermometer. If the junctions are not identical,
the equations become more complicated.
Here, we will consider only the high temperature limit. In this limit
the function $G(t)$ decays at short times of order $\hbar/k_BT$, and one
can
replace the response function $K_j(t)$ in the integral (\ref{IV1}) by
its
short time expansion, which starts from the linear-in-time term,
$K_j(t)\to
\dot K_j(0)t$. Then we get
\begin{eqnarray}
I=\frac{V}{R_\Sigma}+\frac{2e}{\hbar R_\Sigma}\sum\limits_{j=1}^N
\dot K_j(0)
\int\limits_0^{\infty}dt\; G(t)\; t\;
\sin(\frac{eV_jt}{\hbar})\;\;\;\;\;\;\;\;\;\;
 \nonumber\\
=\frac{Nk_BTv}{eR_\Sigma}-\frac{e}{2 R_\Sigma}\sum\limits_{j=1}^N
\dot K_j(0)\left\{\coth\frac{v_j}{2}-\frac{v_j}{2\sinh^2\frac{v_j}{2}}
\right\}.
\label{IV2}
\end{eqnarray}
\begin{equation}
R_\Sigma\frac{dI}{dV}=
1-\sum\limits_{j=1}^N \frac{e^2R_j\dot K_j(0)}{k_BTR_\Sigma}\;
\frac{\frac{v_j}{2}\sinh
v_j-4\sinh^2\frac{v_j}{2}}{8\sinh^4\frac{v_j}{2}}.
\label{G1}
\end{equation}
Here $V_j\equiv VR_j/R_\Sigma$ and $v_j\equiv eV_j/k_BT$. From Eqs.
(\ref{K}) and (\ref{Z})
we find \begin{equation}
\dot K_j(0)=\lim\limits_{\omega\to\infty} (-i\omega Z_j(\omega))=
\frac{1}{C_j}-\frac{1}{C_j^2\sum\limits_{k=1}^N \frac{1}{C_k}}.
\end{equation}
Assuming that deviations, $\delta R_j$ and $\delta C_j$, of the junction
parameters from the reference values $R$ and $C$ are small, we expand
Eq. (\ref{G1}) in powers of $\delta R_j$ and $\delta C_j$
up to the second order. Then we get
\begin{eqnarray}
R_\Sigma \frac{dI}{dV}&=&1-\frac{N-1}{N}\frac{e^2}{Ck_BT}f(v)
-\frac{e^2}{Nk_BT}f(v)\sum\limits_j \delta \dot K_j(0)
\nonumber\\
&&
-\frac{N-1}{N}\frac{e^2}{Ck_BT}[f(v)+vf'(v)]\sum\limits_j\delta r_j
\nonumber\\
&&
-(N-1)\frac{e^2}{Ck_BT}\bigg[vf'(v)+\frac{v^2}{2}f''(v)\bigg]\sum\limits_j
\delta
r_j^2
\nonumber\\
&&
-\frac{e^2}{k_BT}[f(v)+vf'(v)]\sum\limits_j\delta r_j \delta\dot K_j(0).
\label{correction}
\end{eqnarray}
Here
\begin{eqnarray}
f(v)&\equiv &\frac{\frac{v}{2}\sinh
v-4\sinh^2\frac{v}{2}}{8\sinh^4\frac{v}{2}},
\nonumber
\\
\delta r_j&\equiv&\delta\left[\frac{R_j}{R_\Sigma}\right]=
\frac{N\delta R_j-\sum\limits_k\delta R_k}{N^2R}
\left[1-\frac{\sum\limits_k\delta R_k}{NR}\right],
\nonumber
\end{eqnarray}
and
\begin{eqnarray}
\delta\dot K_j(0)&=& -\frac{N-2}{N}\frac{\delta
C_j}{C^2}-\frac{\sum\limits_k \delta C_k}{N^2C^2}
+\frac{N-2}{N}\frac{\delta C_j^2}{C^3}
\nonumber\\
&&
-\frac{\left[\sum\limits_k\delta C_k-N\delta C_j\right]^2}
{N^3C^3}
+\frac{\sum\limits_k\delta C_k^2}{N^2C^3}.
\end{eqnarray}

We observe that $\delta r_j$ is zero if the deviations of all the
resistances are equal to each other. We also note that the terms linear in
$\delta R_j$ vanish in the sum $\sum_j\delta r_j$ and, as a consequence, in Eq.
(\ref{correction}). Both these properties reflect the fact that the
half-width depends only on temperature if the resistances of all the
junctions are the same.
Now the correction to the half-width of the conductance dip
can be obtained peturbatively in $\delta R_j$ and $\delta C_j$.
In the first non-vanishing order we get
\begin{eqnarray}
\frac{V_{1/2}}{V_{1/2,0}}&=&1+\frac{N-2}{N(N-1)}\bigg\{\sum\limits_{j}
\frac{\delta R_j\delta C_j}{RC}
 - \sum\limits_{i,j} \frac{\delta R_i\delta C_j}{NRC}
 \bigg\}
\nonumber\\ &&
-\frac{\alpha}{N}\bigg\{\sum\limits_{j}\frac{\delta
R_j^2}{R^2}-\frac{1}{N}
\bigg[\sum\limits_{j}\frac{\delta R_k}{R} \bigg]^2\bigg\},
\end{eqnarray}
where $\alpha\equiv 1+\frac{v_0}{4}\frac{f''(v_0/2)}{f'(v_0/2)}\approx 0.734$.

\section{Experiment}

To obtain suitable data for comparison between the theoretical predictions
(presented in the previous section) and the experiment, we fabricated high conductance
Al/AlO${_x}$/Al tunnel junction arrays by  electron beam lithography and two-angle shadow evaporation
techniques. As a  substrate, we used nitridized
silicon wafers. The number of junctions in the array, $N$, was twenty, and
each junction had an area of about 0.025 {$\mu$}m{$^2$}. Different high
conductance samples with per-junction asymptotic resistances of
$1-2$ k$\Omega$, together with two samples with lower conductances (in the intermediate regime) with
per-junction resistances equal to 20 k$\Omega$ and 23 k$\Omega$, were made and measured.
In order to decrease heating effects in the high conductance
arrays at higher bias voltages, the islands between the junctions in most of the samples
(albeit not for those shown in Fig. 3) were made
sufficiently large with cooling bars (see, {\it e.g.}, \cite{exp4})
attached to them. The measurements were carried out in the temperature range
1.5 K $\lesssim T\lesssim$ 4.5 K, {\it i.e.} at temperatures of liquid helium. To
measure the temperatures as accurately as possible,
we fabricated CBT sensors on the same sample stage in the vicinity
of the samples to be measured.

As it was already discussed above, Coulomb blockade -- although
weakened -- is not smeared out completely even in the strong
tunneling limit. The zero bias conductance of the array is always
lower than its asymptotic value at high voltages. In general, the
zero bias conductance at high temperatures can be written as

\begin{equation}
\frac{G_0}{G_\Sigma}=1-\frac{N-1}{N}\frac{E_C}{3k_BT}+
A\frac{E_C^2}{(k_BT)^2}+\cdots .
\label{highT}
\end{equation}
In the limit of strong tunneling, $R_j\ll R_{\rm q}$, from Eq.  (\ref{Rsigma}) we find
$
A= A_{strong}\equiv\frac{N-1}{N}\frac{3\zeta(3)}{2\pi^4}\frac{R_{\rm
q}}{R}=0.0185\frac{N-1}{N}\frac{R_{\rm q}}{R},
$
while in the opposite weak tunneling limit the approach based on
the Master equations, \cite{FHK}, yields
$
A=A_{weak}\equiv\frac{1}{15}\left(\frac{N-1}{N}\right)^2.
$
Here, as before, $R\equiv{R_j}$ is the (per-junction) resistance of the
homogeneous array at large bias voltages.

\begin{figure}
\begin{center}
\includegraphics[height=6.5cm]{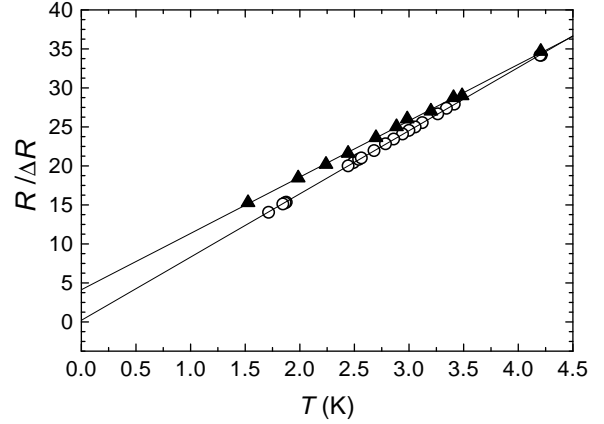}
\end{center}
\caption{The temperature dependence of the zero bias voltage anomaly, $R/\Delta R$, for two arrays with
per-junction resistances of 1.2 k$\Omega$ and 23 k$\Omega$ (solid triangles and open circuits, respectively).
The solid lines are linear fits to the experimental data.} 

\end{figure}

In the intermediate regime, which is appropriate for most of the arrays
measured in the experiment, one can conjecture that $A=A_{strong}+A_{weak}$.
Previously, this conjecture was verified for the specific case of SET
transistors ($N=2$) \cite{G,GG}.

Within the first order in $u=\hbar/{2\pi{k_B} TRC}$, the inverse resistance enhancement at zero bias voltage,
$R/\Delta R$, where $\Delta R\equiv R(V=0)-R$, can be easily derived from Eq. (\ref{highT}) as:

\begin{equation}
\frac{R}{\Delta R}= 3\frac{N}{N-1} \frac{k_B }{E_C}T +
a\frac{R_{\rm q}}{R}+b,
\label{DeltaR}
\end{equation}
where, according to the theory,
\begin{equation}
a=\frac{N}{N-1}\frac{27\zeta(3)}{2\pi^4}=0.175
\label{a}
\end{equation}
and
\begin{eqnarray}
b=\cases{-1, &$A=A_{strong};$\cr
-2/5, &$A=A_{strong}+A_{weak}.$\cr}
\label{b}
\end{eqnarray}

The expression above has two characteristic features: linearity in $T$ and dependence of its slope on the
capacitance of the junctions in the array. In addition, this equation predicts an offset which depends {\it only} on
the number of junctions in the array, and on the ratio of the quantum and per-junction resistances. Below, while
making comparison between the measured data and the predictions of Eq. (\ref{DeltaR}), we will take the
the weak tunneling correction into account, {\it i.e}, $A=A_{strong}+A_{weak}$.

The value $R/\Delta R$,  measured for two tunnel junction arrays
with $N=20$ and at different temperatures, is displayed in Fig. 3.
The asymptotic resistances of the samples were 23 k$\Omega$ (open
circles) and 1.2 k$\Omega$ (solid triangles).  One observes an
almost perfect linear dependence of the value $R/ \Delta R$ on
temperature (cf. Eq. (\ref{DeltaR}). This dependence can be used
to obtain a quantitative estimate for the junction capacitance. By
fitting the slope of the experimental curves of Fig. 3 to Eq.
(\ref{DeltaR}) we find $C=2.4$ fF and 2.1 fF respectively for the
arrays with  $R=23$ and 1.2 K$\Omega$. This way appears to be the
most reliable to evaluate the per-junction capacitance in arrays
of normal metal tunnel junctions \cite{exp1,exp2}.

The second and third terms in the right hand side of Eq. (\ref{DeltaR})
yield an offset which -- for a given number of
junctions $N$ -- should depend solely on the ratio $R_{\rm q}/R$.
Eq. (\ref{DeltaR}) gives the values  3.4 and -0.2 for this offset respectively for the
samples represented by the solid triangles and the open circles. The corresponding numbers
obtained from the lines fitted to the experimental data in Fig. 3 are 4.1 and 0.2, being  in a reasonable
agreement with the above theoretical values.

\begin{figure}
\begin{center}
\includegraphics[height=6.5cm]{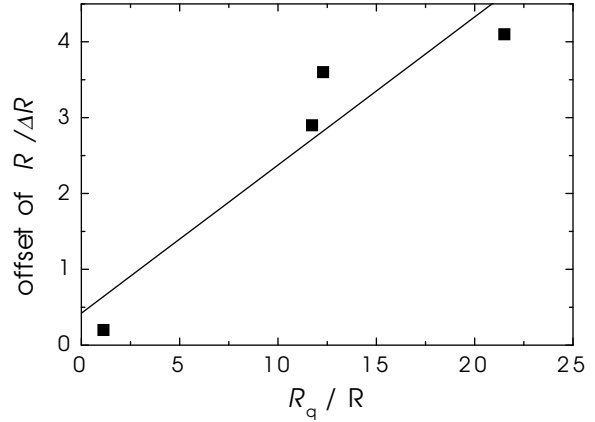}
\end{center}
\caption{The dependence of the offset of $R/\Delta R$ (cf. Fig. 3)
on the dimensionless resistance, $R_{\rm q}/R$, of the junctions. The measured data are shown
by  squares, the solid line is the best linear fit.}
\end{figure}

The offset values for two additional samples with (per-junction) resistances $R=2.1$ k$\Omega$ and
$R=2.2$ k$\Omega$ (the samples for which the systematic $T$-dependence was measured) were equal to 3.6
and 2.9, respectively. The corresponding theoretical predictions are 1.8 and 1.2.  The data for the offsets are
presented in Fig. 4. We observe that arrays with lower resistances show larger offsets, as predicted by our 
theory,
Eq. (\ref{DeltaR}). However, the offset values presented as a function of
$R_{\rm q}/R$ do not exactly fall on the straight line: they are scattered
within an interval of $\pm 0.5$. This effect could probably be
attributed to the inhomogeneity of the arrays. For instance, the measured offsets for two arrays
with nearly identical values of $R=2.1$ k$\Omega$ and 2.2 k$\Omega$ differ
from each other. This discrepancy can be explained, if one assumes
a certain degree of  junction asymmetry, {\it e.g.}, a 20 $\%$ fluctuation of the
per-junction resistance, $R_j$, around the mean  value, $R$. This, in turn, can arise from a 40 \%
fluctuation in the area of the "odd" and the "even" tunnel junctions as a consequence
of the two-angle evaporation technique employed in the sample fabrication.

The best linear fit for the offset -- inverse resistance
dependence (Fig. 4) yields $a=0.19\pm 0.05$, $b=0.4\pm 0.7$. The
value $a$ agrees well with our theoretical prediction (\ref{a}).
At the same time the experimental value of $b$ turns out to be
different from the predictions of the theory (\ref{b}). It is also
too uncertain due to the scattering of the data points. At
present, possible reasons for this discrepancy remain unclear. We
have checked asymmetry effects as well as those of the external
environment within the framework of the Eq. (\ref{dIdV}). These
effects can hardly help to improve the agreement between
theoretical and experimental values of $b$.


In the weak tunneling regime $R\gg R_{\rm q}$ the offset is given
by $b$ because $aR_{\rm q}/R\to 0$. In order to check the effect
of finite external resistance on the offset in this limit we have
performed Monte-Carlo simulations (for details see
\cite{exp3,HPP}) based on the phase-correlation theory \cite{IN}.
The only fitting parameter used in these simulations was the
junction capacitance. By comparing the measured conductance curves
of the array represented by the solid triangles in Fig. 3 ($R=2.1$
k$\Omega$) to those obtained from the numerical simulations, we
have obtained the value $C=2.1$ fF for this sample. Notice that
this value is in excellent agreement with that derived from 
Eq. (\ref{DeltaR}) as explained above. We have also found that,
depending on the magnitude of the environmental resistance, $b$
may increase by $0.1$ at most. Again this is not sufficient to
explain the observed values of $b$.


Figure 5 shows the variation of the normalized half-width
$\Delta V_{1/2}/V_{1/2,0}$
as a function of temperature ($\Delta V_{1/2}\equiv V_{1/2}-V_{1/2,0})$. The solid and open squares
represent the data
measured for two 20-junction arrays both with $R=2.1$ k$\Omega$.
The solid curve corresponds to our theoretical
prediction, Eq. (\ref{Vhalfexact}). We observe, that the difference between the
experimental points and our theoretical curve typically does not exceed
one percent, {\it i.e.} the ageement is fairly good.

\begin{figure}
\begin{center}
\includegraphics[height=6.5cm]{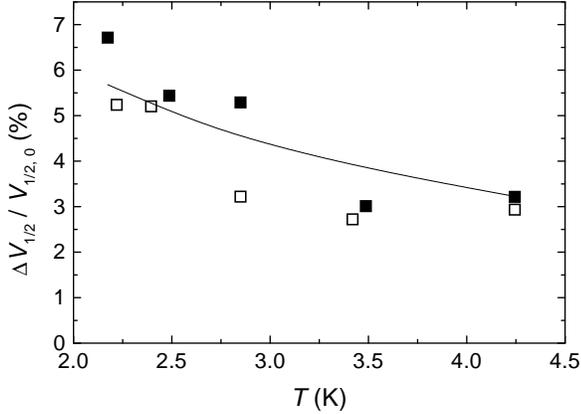}
\end{center}
\caption{The measured temperature dependence of the normalized half-width
$\Delta V_{1/2}/V_{1/2,0}$ for two arrays (solid and open squares)
with the same per-junction resistances $R = 2.1$ k$\Omega$.
The solid curve has been obtained from the strong tunneling theory (see the text).}
\end{figure}

Within the framework of our analysis, one can easily establish yet one more useful
relation between the normalized half-width,
$\Delta V_{1/2}/V_{1/2,0}$, and the value of
the conductance dip $\Delta G \equiv G_\Sigma -G(V=0)$
[cf, {\it e.g.},  Eq. (\ref{dIdV})] evaluated in the linear regime.
Combining Eqs. (\ref{Vhalfexact}) and (\ref{Rsigma}) in the high
temperature limit one gets
\begin{equation}
\frac{\Delta V_{1/2}} {V_{1/2,0}}= \chi(R_{\rm q}/R)\frac{\Delta
G}{G_\Sigma},
\label{strongcorr}
\end{equation}
where the function $\chi(R_{\rm q}/R)$ in the strong tunneling limit
$R_{\rm q} \gg R$ reads
\begin{equation}
\chi (R_{\rm q}/R) \simeq 0.108 \frac{N}{N-1}\frac{R_{\rm q}}{R}.
\label{chi1}
\end{equation}
Note, that Eq. (\ref{strongcorr}) was previously derived
in Ref. \cite{exp1} in the opposite weak tunneling limit $R \gg R_{\rm q}$,
in which case the $\chi$-function tends to the constant
\begin{equation}
\chi (R_{\rm q}/R) \simeq 0.392.
\label{chi2}
\end{equation}
We assume that in the intermediate regime the function $\chi(R_{\rm q}/R)$
is the sum of the expressions (\ref{chi1}) and (\ref{chi2}). This assumption,
although not proven rigorously, provides a reasonable interpolation between
the two limiting cases.

In addition to the arrays discussed above, three other samples with
per-junction resistances of 20 k$\Omega$, 1.4 k$\Omega$, and 1.0 k$\Omega$ were measured at
$T\simeq 4.2$ K. The exact temperature was obtained from the vapor pressure of liquid helium.
The corresponding data are collected into Table 1. The second row in the table represents the
measured half-widths obtained {\sl directly} from the measurement, $V_{1/2, \rm{meas}}$.
The corresponding conductance dips, $\Delta G/G_\Sigma$,
are shown in the third row. The forth row demonstrates the ''corrected'' half-widths
(i.e. the weak tunneling correction  [Eqs. (\ref{strongcorr}) and (\ref{chi2}) subtracted],
\begin{equation}
V_{1/2,{\rm corr.}}=V_{1/2,{\rm meas.}}[1-0.392\frac{\Delta G}{G_\Sigma}].
\end{equation}
In the weak tunneling regime this value would coincide with $V_{1/2,0}=5.439 Nk_BT/e.$
However in our experiments this is not the case.
The fifth row shows the
relative deviations of the corrected half-widths from that of the basic linear result,
$\Delta V_{1/2,0}^{\rm corr}=(V_{1/2,{\rm corr}}-V_{1/2,0})/V_{1/2,0}$. In
fact, the numbers in this row can be interpreted as the "residual inaccuracies" of the measured half-widths as
compared to the weak tunneling approximation. They can be explained by the strong
tunneling effects.
The corresponding theoretical values, obtained from the strong
tunneling correction to the half-width of the conductance dip
around zero bias voltage [Eq. (\ref{strongcorr}, \ref{chi1})], are shown in the last row of the table. These 
predictions are
in a good agreement with the corrected measured data discussed above.

\begin{table}\centering
\begin{tabular}{|l|l|l|l|l|l|}
\hline
\vbox to3.2ex{\vspace{1pt}\vfil\hbox to18ex{\hfil $R$ (k$\Omega$)\hfil}} &
\vbox to3.2ex{\vspace{1pt}\vfil\hbox to4.6ex{\hfil 20\hfil}} &
\vbox to3.2ex{\vspace{1pt}\vfil\hbox to4.6ex{\hfil 2.1\hfil}} &
\vbox to3.2ex{\vspace{1pt}\vfil\hbox to4.6ex{\hfil 2.1\hfil}} &
\vbox to3.2ex{\vspace{1pt}\vfil\hbox to4.6ex{\hfil 1.4\hfil}} &
\vbox to3.2ex{\vspace{1pt}\vfil\hbox to4.6ex{\hfil 1.0\hfil}} \\
\hline
\vbox to3.2ex{\vspace{1pt}\vfil\hbox to18ex{\hfil $V_{1/2,{\rm meas.}}$ (mV)\hfil}} &
\vbox to3.2ex{\vspace{1pt}\vfil\hbox to4.6ex{\hfil 40.20\hfil}} &
\vbox to3.2ex{\vspace{1pt}\vfil\hbox to4.6ex{\hfil 41.90\hfil}} &
\vbox to3.2ex{\vspace{1pt}\vfil\hbox to4.6ex{\hfil 41.32\hfil}} &
\vbox to3.2ex{\vspace{1pt}\vfil\hbox to4.6ex{\hfil 42.22\hfil}} &
\vbox to3.2ex{\vspace{1pt}\vfil\hbox to4.6ex{\hfil 41.91\hfil}} \\
\hline
\vbox to3.2ex{\vspace{1pt}\vfil\hbox to18ex{\hfil $\Delta G/G_{\Sigma} (\%)$\hfil}} &
\vbox to3.2ex{\vspace{1pt}\vfil\hbox to4.6ex{\hfil 2.15\hfil}} &
\vbox to3.2ex{\vspace{1pt}\vfil\hbox to4.6ex{\hfil 2.32\hfil}} &
\vbox to3.2ex{\vspace{1pt}\vfil\hbox to4.6ex{\hfil 2.26\hfil}} &
\vbox to3.2ex{\vspace{1pt}\vfil\hbox to4.6ex{\hfil 2.21\hfil}} &
\vbox to3.2ex{\vspace{1pt}\vfil\hbox to4.6ex{\hfil 1.79\hfil}} \\
\hline
\vbox to3.2ex{\vspace{1pt}\vfil\hbox to18ex{\hfil $V_{1/2,{\rm corr.}}$ (mV)\hfil}} &
\vbox to3.2ex{\vspace{1pt}\vfil\hbox to4.6ex{\hfil 39.85\hfil}} &
\vbox to3.2ex{\vspace{1pt}\vfil\hbox to4.6ex{\hfil 41.52\hfil}} &
\vbox to3.2ex{\vspace{1pt}\vfil\hbox to4.6ex{\hfil 40.96\hfil}} &
\vbox to3.2ex{\vspace{1pt}\vfil\hbox to4.6ex{\hfil 41.86\hfil}} &
\vbox to3.2ex{\vspace{1pt}\vfil\hbox to4.6ex{\hfil 41.62\hfil}} \\
\hline
\vbox to3.2ex{\vspace{1pt}\vfil\hbox to18ex{\hfil $\Delta V_{1/2}^{\rm corr.}/V_{1/2,0} (\%) $\hfil}} &
\vbox to3.2ex{\vspace{1pt}\vfil\hbox to4.6ex{\hfil 0.5\hfil}} &
\vbox to3.2ex{\vspace{1pt}\vfil\hbox to4.6ex{\hfil 4.7\hfil}} &
\vbox to3.2ex{\vspace{1pt}\vfil\hbox to4.6ex{\hfil 3.3\hfil}} &
\vbox to3.2ex{\vspace{1pt}\vfil\hbox to4.6ex{\hfil 5.6\hfil}} &
\vbox to3.2ex{\vspace{1pt}\vfil\hbox to4.6ex{\hfil 5.0\hfil}} \\
\hline
\vbox to3.2ex{\vspace{1pt}\vfil\hbox to18ex{\hfil $\Delta V_{1/2}^{\rm theor.}/V_{1/2,0} (\%) $\hfil}} &
\vbox to3.2ex{\vspace{1pt}\vfil\hbox to4.6ex{\hfil 0.3\hfil}} &
\vbox to3.2ex{\vspace{1pt}\vfil\hbox to4.6ex{\hfil 3.2\hfil}} &
\vbox to3.2ex{\vspace{1pt}\vfil\hbox to4.6ex{\hfil 3.2\hfil}} &
\vbox to3.2ex{\vspace{1pt}\vfil\hbox to4.6ex{\hfil 4.8\hfil}} &
\vbox to3.2ex{\vspace{1pt}\vfil\hbox to4.6ex{\hfil 5.3\hfil}} \\
\hline
\end{tabular}
\caption{Data for different samples at $T\simeq4.2$ K. Samples
with $R=2.1$ k$\Omega$ were further measured at lower temperatures
(Fig. 5). The fifth row shows the measured data, whereas the last
row has been obtained from the theory [Eqs. (36) and (37)]. Here
we defined $\Delta V_{1/2}^{\rm corr.}\equiv V_{1/2,
\rm{corr}}-V_{1/2, 0}$ (see also the text).}
\end{table}
\section{Conclusions}

We have studied one-dimensional arrays of tunnel junctions in the strong
tunneling regime, both theoretically and  experimentally. Within
the framework of the quasiclassical Langevin equation formalism, analytical
expressions for the current-voltage characteristics of such arrays,
together with the expressions for the half-width of the conductance dip
around zero bias voltage, have been derived. Furthermore, the effect of external
elecromagnetic environment has been studied theoretically. We have
fabricated and measured several arrays of tunnel junctions with
per-junction resistances ranging from 1 k$\Omega$ to 23 k$\Omega$. Our measured data
are in rather good agreement with the theoretical predictions.
The experiments demonstrate that Coulomb blockade effects survive even in
the strong tunneling regime, and are
clearly visible in arrays with per-junction tunnel resistances as low as 1 k$\Omega$. These
observations are in agreement with other recent experimental results on SET
transistors \cite{Chalmers} and  on single junctions \cite{Saclay}-\cite{Helsinki}.

It has been verified, both in the theory and in the experiment,
that high conductance arrays of tunnel junction are less favorable
for Coulomb Blockade Thermometry (CBT) applications. This is a
combined consequence of the more pronounced heating effects in
such arrays, {\it and} the larger departure from the simple linear
relation $V_{1/2,0}=5.439Nk_BT/e$, which is of central role in the
CBT applications. The latter, however, is not such a severe
limitation, since corrections to the linear relation are now well
known for all values of the junction resistance. It has also been
shown that in the high temperature limit (in the leading
approximation in $1/T$) the results for the conductance dip, now
derived for arbitrary tunneling strength, coincide with those
previously obtained in the weak tunneling regime.



\end{document}